\documentclass[prb,aps,epsfig,twocolumn,showpacs]{revtex4-1}
\usepackage{color}
\usepackage{cancel}
\usepackage{lipsum}
\usepackage{epsfig,amssymb,amsmath,latexsym}
\usepackage{float}
\usepackage{color}
\usepackage[colorlinks=true,linktoc=page,linkcolor=magenta,citecolor=magenta]{hyperref}
\begin{document}
	\textheight=23.8cm

\title{ Driven quantum spin chain in the presence of noise: Anti-Kibble-Zurek behavior}

\author{Manvendra Singh, and Suhas Gangadharaiah}
\affiliation {Department of Physics, Indian Institute of Science Education and Research, Bhopal, India}

\date{\today}

\begin{abstract}
We study defect generation in a quantum  XY-spin chain arising due to the linear drive of the many-body Hamiltonian in the presence of a time-dependent fast Gaussian noise. The main objective
 of this work is to quantify analytically the effects of noise on the defect density production. In the absence of noise, it is well known that in the slow sweep regime, the defect density follows the Kibble-Zurek (KZ) scaling behavior with respect to the sweep speed. We consider time-dependent fast Gaussian noise in the anisotropy of the spin-coupling term [$\gamma_0=(J_1-J_2)/(J_1+J_2)$] and show via analytical calculations that the defect density exhibits anti-Kibble-Zurek (AKZ) scaling behavior in the slow sweep regime. 
In the limit of large chain length and long time, we calculate the entropy and magnetization density of the final decohered state  and show that their scaling behavior is consistent with AKZ picture in the slow sweep regime.  
 We have also numerically calculated the sub-lattice spin correlators for finite separation by evaluating the Toeplitz determinants and find results consistent with the KZ picture in the absence of noise, while in the presence of noise and slow sweep speeds the correlators exhibit the AKZ behavior.
  Furthermore, by considering the large $n$-separation asymptotes of the Toeplitz determinants, we further quantify the effect of the noise on the  spin-spin correlators in the final decohered state. We show that while the correlation length of the sub-lattice correlator scales according to the AKZ behavior, we obtain different scaling for the magnetization correlators. 
 \end{abstract}


\maketitle

\section{Introduction}
A quantum system driven at  zero temperature by some  system dependent parameter through a quantum critical point (QCP)
 is subjected to quantum phase transition (QPT) in which
the ground state of the system is fundamentally altered
 with completely different physical properties across the phase transition point.  
 One of the main points of interest is to 
quantify the generation of excitations or the defect density generation due to the quench
through the critical point.
Defects are inevitable in a drive through the critical point due to the
vanishing energy gap at the critical point where the adiabaticity criterion breaks down and non-adiabatic effects become important. In this regard, the Kibble-Zurek mechanism (KZM) a theory originally proposed to quantify the topological defect production in a cosmological phase transition has been successfully applied in quantifying the defect production in idealized condensed matter systems  
undergoing QPT ~\cite{Kibble_1976, KIBBLE1980183, Zurek1985, Kibble1985, Zurek1994, ZUREK1996177, doi:10.1142/S0217751X1430018X, PhysRevLett.95.035701, PhysRevLett.95.105701, PhysRevLett.95.245701, PhysRevB.72.161201, RevModPhys.83.863}. The theory predicts that the defect density scales as $n\propto \tau^{-\beta}$, where $\tau$ is the quench rate and the universal exponent  $\beta>0$ is determined by the critical exponents and the dimension of the system. 
 Recent experimental studies supporting KZM
have been reported in well controlled systems involving trapped ions, Bose-Einstein condensates and Rydberg simulator~\cite{Lamporesi2013, Cui2016, Keesling2019}.

While the study of quantum systems exhibiting KZ behavior 
remains an area of active interest, scenarios which result 
in deviations from this universal behavior have also come 
under increased scrutiny.  Recent studies of drive 
protocols in quantum systems that are  coupled to external 
environment, disorder or are in the presence of noise indicate that the defect density generated 
exhibit fundamentally  different dynamical behavior 
than the one predicted by the KZ theory~\cite{PhysRevB.74.064416, PhysRevA.76.030304, Fubini_2007,PhysRevLett.101.175701,  PhysRevB.80.024302, PhysRevA.82.013630, PhysRevLett.107.137201,PhysRevLett.106.217206, PhysRevLett.109.045302,PhysRevB.86.060408,PhysRevLett.111.150403, PhysRevB.87.174301,PhysRevB.89.024303, PhysRevA.90.062130, Zueco_2008,PhysRevA.97.033626, PhysRevB.101.104307, PhysRevLett.124.090502}. The focus of our attention has been to understand  experimental and numerical studies wherein, unlike the KZ behavior, slower drives beyond a certain optimal quench rate/speed  create more defects~\cite{PhysRevX.2.041022, YUKALOV20151366, PhysRevLett.117.080402, PhysRevB.95.224303, PhysRevX.7.041014, Gardas2018, PhysRevLett.124.230602}. 
This scaling has  been termed as the anti-Kibble-Zurek (AKZ) behavior. In all of them, the AKZ
scaling behavior manifested itself in the presence of the noisy 
control field driving the system through the critical points. 
In this work, we consider a time dependent quantum XY-spin chain which is driven by a transverse magnetic field  with a anisotropy term that contains a  fluctuating Gaussian noise term. In the limit of fast noise, we perform exact analytical calculations and derive the universal AKZ scaling behavior of the defect density with respect to the quench rate. 

Apart from the study of the defect generation, 
the consequence of the KZ picture to the entropy, magnetization and the correlation functions have been considered before (for example in the~Refs.~[\onlinecite{PhysRevA.73.043614, PhysRevA.75.052321, PhysRevLett.100.077204, PhysRevB.78.045101, PhysRevB.79.045128, PhysRevB.89.024303}]).  We furthermore  quantify the effect of noise on the above physical quantities in the final decohered state due to the noisy drive through the QCPs. By considering the large $n$-separation asymptotes of the Toeplitz determinants, we show analytically that the  correlation lengths of the sub-lattice correlators exhibits the AKZ scaling behavior in the slow sweep  regime. The scaling behavior of the magnetization density at the end of the protocol is also consistent with the AKZ scaling behavior in the slow sweep limit. However,  the correlation length of the (connected) magnetization correlator 
in the large $n$-separation limit continues to follow the KZ picture. 

The organization of the paper is as follows. In Sec.~II we discuss the model Hamiltonian and dynamics of the  XY-spin chain with transverse magnetic field (varying linearly with time) in the $z$-direction in the absence of noise.  
In Sec.~III we consider transverse protocol in the presence of fast Gaussian noise and obtain analytically the AKZ scaling behavior of the defect density in the slow drive regime. 
We also derive an expression for the optimal quench time
with which the system must be driven so as to minimize the defect production at the end of the drive protocol. 
In Sec.~IV we discuss the decoherence of the local observables due to the drive  through the QCPs in the presence of Gaussian noise and in addition the expectation values of a fermionic $2$-point correlator in the final decohered state has been obtained. In Sec.~V we derive the analytical expression for the entropy density for the final decohered state and show that the results are consistent with the 
AKZ scenario. 
Finally, in Sec.~VI we discuss in detail the analytical results  for the spin correlators and the magnetization density at the end of the noisy drive protocol. We have summarized our results in Sec.~VII.

\section{The Model Hamiltonian}
We consider the quantum XY-spin model driven by the transverse external field $h(t)$,
\begin{equation}
    H(t)=- \sum^{N}_{n=1}\left[ J_1 \sigma^{1}_{n}\sigma^{1}_{n+1}+J_2 \sigma^{2}_{n}\sigma^{2}_{n+1}+ h(t) \sigma^{3}_{n}\right],
\end{equation}
where, $J_1$ and $J_2$ are respectively the spin-spin couplings along the $x$ and $y$-spin directions. Introducing the coefficients $J=J_1 +J_2$ and anisotropy, $\gamma_0=(J_1-J_2)/J$, allows us to re-express the Hamiltonian as,
\begin{align}
  H(t)= -\frac{J}{2}\sum^{N}_{n=1}   \left[(1+\gamma_0) \sigma^{1}_{n}\sigma^{1}_{n+1}+ (1-\gamma_0) \sigma^{2}_{n}\sigma^{2}_{n+1} \right]\\ \notag - h(t)\sum^{N}_{n=1}  \sigma^{3}_{n} .
  \label{eq:Hamiltonian2}
\end{align}
 The $\gamma_0=0 $ limit represents the isotropic XY-spin chain, while $\gamma_0=\pm 1$ limits correspond to the quantum Ising chain case. We consider time dependence in the anisotropy by including Gaussian correlated noise $\eta(t)$, i.e., $\gamma_0\rightarrow \gamma(t)=\gamma_0+\eta(t)$. The Gaussian noise $\eta(t)$ is characterized by,
\begin{equation}
\overline{\eta(t)}=0, \,\, \overline{\eta(t)\,\eta(t_1)}=\eta^2_0 \, e^{-\Gamma|t-t_1|},
\end{equation} 
where $\eta_0$ is the noise strength and $\Gamma$ is the inverse time-scale associated with the noise.

In the following, we will summarize the well studied transverse protocol~\cite{PhysRevA.73.043614, Divakaran_2009, PhysRevB.78.144301}. In this protocol, the transverse external field $h(t)$ is tuned to drive the equilibrium system from  $h\rightarrow -\infty$ at the start of the protocol to $h\rightarrow \infty$ at the end of the drive protocol. For a linear protocol, 
$$h(t)= v t= J t/\tau_Q,$$
where $v$ is the sweep speed of the drive,  $\tau_Q=J/v$ is the `quench time' and the time $t$ runs from $-\infty$ to $\infty$. The system starts out in the paramagnetic (PM) ground state (GS), $|\downarrow \downarrow \downarrow...\downarrow\rangle$  with all the spins along the negative $z$-axis. As $t$ is increased the system goes through the QCPs at $h(t)=\pm J$. The quantum phase transition involves change in the nature of GS from PM to ferromagnetic (FM) at $h=-J$ and from FM to PM at $h=J$.
The final state that emerges is not the perfect GS of the Hamiltonian, $|\uparrow \uparrow \uparrow  ...\uparrow\rangle$, but instead is formed out of the quantum superposition of the states of the type $|..\uparrow \uparrow  \downarrow  \uparrow \uparrow \uparrow  \uparrow  \downarrow \uparrow \uparrow..\rangle$.
The reason for such a final state is the inevitable violation of the adiabaticity criteria.  
This criteria is obtained by comparing two time-scales, the relaxation time scale, which is proportional to the inverse of minimum gap of the system and a time scale for driving the system, $\tau_Q $. For a perfectly adiabatic dynamics, `$\mathrm{relaxation\, time} \ll \tau_Q$'. But in the $N\rightarrow \infty$ limit, the energy gap vanishes at the QCPs ($h(t)=\pm J$). Therefore the dynamics becomes non-adiabatic in close proximity to the QCPs, leading to a final state which is formed out of the 
quantum superposition of the states having kinks/domain walls. 

It turns out that rate of production of these topological defects can be quantified with the help of KZ scaling theory. 
Qualitatively one can understand the theory as follows~\cite{dutta_aeppli_chakrabarti_divakaran_rosenbaum_sen_2015}. The energy gap around the QCP depends on the driving field, $\Delta(h)\sim|h-h_c|^{\nu\, z_d}\sim|vt|^{\nu z_d}$ (assuming the gap varies linearly with time) 
with $\nu$ being the correlation length exponent and $z_d$ the dynamical exponent. The correlation length diverges near the QCP as $\xi\sim |h-h_c|^{-\nu}$, whereas the excitation energy, $E(k,h=h_c)\sim |k-k_c|^{z_d}$, characterized by the dynamical exponent $z_d$, vanishes near the critical mode $k_c$.
The region where the adiabaticity breaks down called the non-adiabatic/impulse region (near the QCP) can be estimated by comparing the rate of change of the driving parameter/energy-gap with the energy which is proportional to the square of the energy i.e.,
$
d\Delta/dt\approx \Delta^2.
$
 One finds that the energy scale at which the adiabaticity breaks down is given by
$\Delta^{*}\sim |v|^{\nu\,z_d/(\nu\,z_d+1)}$. Corresponding to this energy scale a length scale $\xi^{*}\sim |v|^{-\nu/(z_d \nu+1)}$ can be associated beyond which the fluctuations of the order parameter cannot follow the adiabatic dynamics resulting in the creation of defects/excitations. 
\begin{equation}
n\sim |\xi^{*}|^{-d}\sim |v|^{\nu d/(\nu z_d+1)}\propto  |\tau_Q|^{-\nu d/(\nu z_d+1)}  
\end{equation} 
where $d$ is the dimension of the system.
For a 1-D XY-spin chain under going a transverse protocol (linearly with time) and $\nu=z_d=1$, the KZ scaling theory predicts the defect density to scale as  $n\sim \sqrt{v}\sim 1/\sqrt{\tau_Q}$.

\begin{figure}
\includegraphics[scale=0.38]{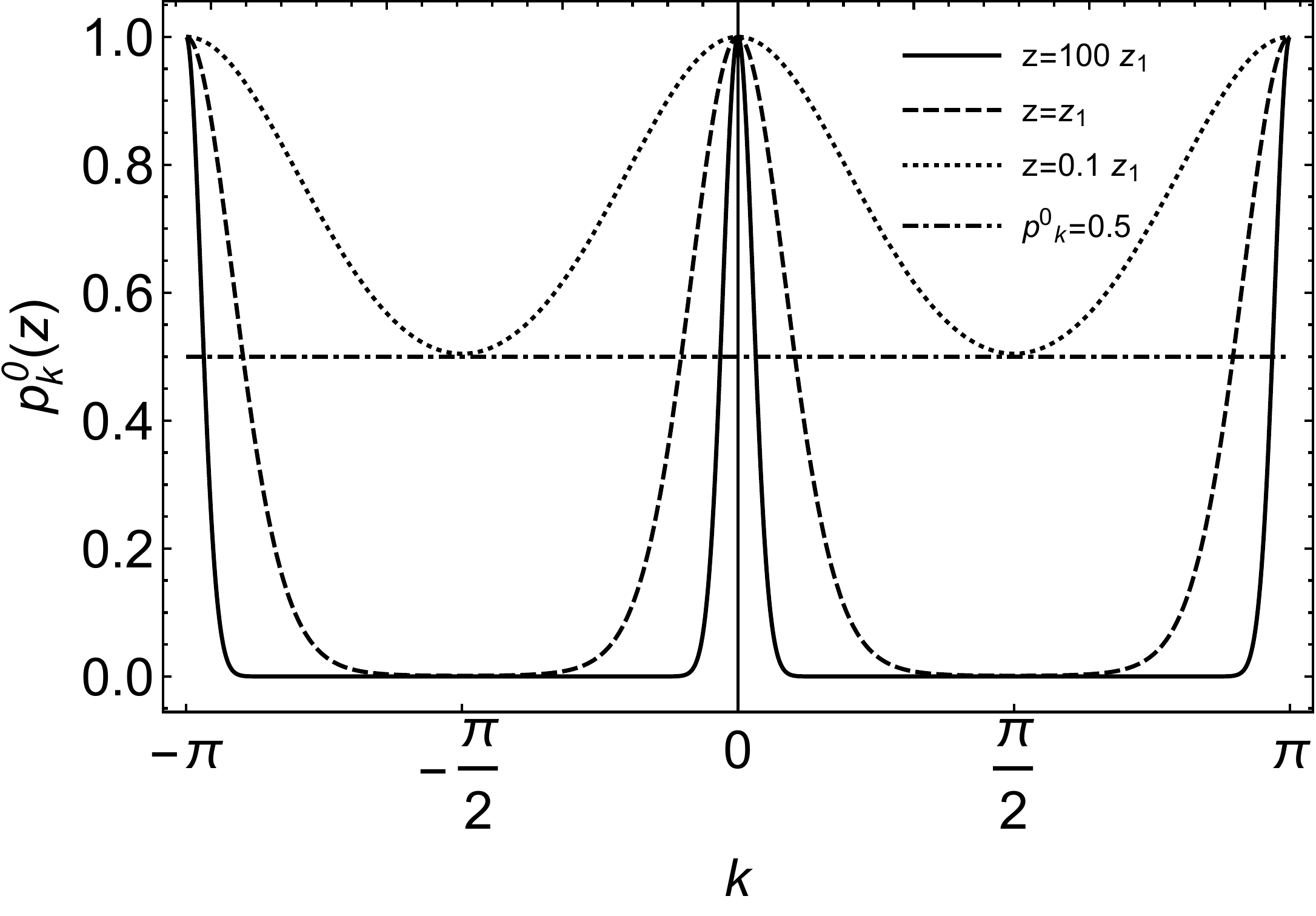} \caption{$p^{0}_k$ Vs $k$ in the absence of noise. In the large $z$ or the small sweep speed limit,  most of the contribution to the defect generation comes from the regions near the critical points $k=0, \,\pm \pi$. It is to be noted that $z_1=\log 2/2\pi$ is a special value of $z$, where $p^{0}_k=1/2$ at $k=\pm \pi/2$. } \label{fig:Pk0}
\end{figure}

These predictions were confirmed by the exact solution of the defect density generation in a linearly driven XY-spin chain Hamiltonian by mapping it to a set of independent two-level Landau-Zener (LZ) problems. 
The first step involves the use of Jordan-Wigner (JW) transformation, $\sigma^{\pm}_n=(e^{\pm i \pi \sum^{n-1}_{l=1}} c^{\dagger}_l c_l) c_n$ and $\sigma^z_l=2  c^{\dagger}_l c_l-1$ (with $\sigma^{\pm}_l=\sigma^x\pm i \sigma^y$), to map the  spin-1/2 Hamiltonian  to the spinless free fermion Hamiltonian 
\begin{eqnarray}
H(t)&=& -J\,\sum^N_{n=1}[( c^{\dagger}_n c_{n+1} +\gamma_0 c_{n+1}  c_n + \text{h.c.}){}\notag\\
&&- h(t) (2c^{\dagger}_n \,c_n-1)  ].\label{eq:H_quadratic}
\end{eqnarray}
Restricting   to the even parity subspace with $c_{N+1}=-c_1$ and using the Fourier transformation, $c_n= \frac{e^{-i\,\pi/4}}{\sqrt{N}} \sum_k c_k\,e^{i\,k\,n} $  yields
\begin{align}
H(t)=\sum_k \,\Big[\left(h(t)+J\cos{k}  \right)c^{\dagger}_k c_k +\frac{\Delta_k}{2} c_{-k} c_k\Big] +\text{h.c.},
\end{align}
where $\Delta_k=2 J\, \gamma_0\sin(k) $.
The above Hamiltonian  can be written as a sum of independent terms, $H(t)=\sum_{k>0}\,H_k (t)$, where for each value of k, $H_k (t)$ acts on the $4$-dimensional Hilbert space spanned by the basis vectors: $|0\rangle,\, |k\rangle=c^{\dagger}_k |0\rangle,\, |-k\rangle=c^{\dagger}_{-k} |0\rangle\, \mathrm{and} \, |k,-k\rangle=c^{\dagger}_k c^{\dagger}_{-k}  |0\rangle$.   
We note that the Hamiltonian with or without the noise term proportional to $\eta(t)$   leaves the parity unchanged.
Although the $1$-particle states $|k\rangle$ and $|-k\rangle$ evolve in time with a global phase factor  only, the states $|0\rangle$ and $|k,-k\rangle$ couple to each other and exhibit LZ dynamics. 
The projected Hamiltonian in the subspace of $|0\rangle$ and $|k,-k\rangle$
has the structure of LZ type Hamiltonian  and is given by
\begin{equation}
\mathcal{H}_k(t)= 2\, \begin{bmatrix}
h(t)+ J\cos(k)& \Delta_k/2  \\
\Delta_k/2 & -h(t)-J\cos(k) 
\end{bmatrix},
\label{eq:H_LZ}
\end{equation}
where note that the off-diagonal term  is  $k$-dependent. 
Applying the LZ transition theory, yields the $k$-dependent transition/excitation probability (in the large time limit),
\begin{equation}
p^0_k(z)=e^{-2\pi z \sin^2{k}},
\label{eq:Pk0}
\end{equation}
where 
$z$ is the dimensionless quench parameter given by $z=J^2 \gamma_0^2/v=J^2 \gamma_0^2 \tau_Q$ (see Fig.~\ref{fig:Pk0}).

\begin{figure}
\includegraphics[scale=0.38]{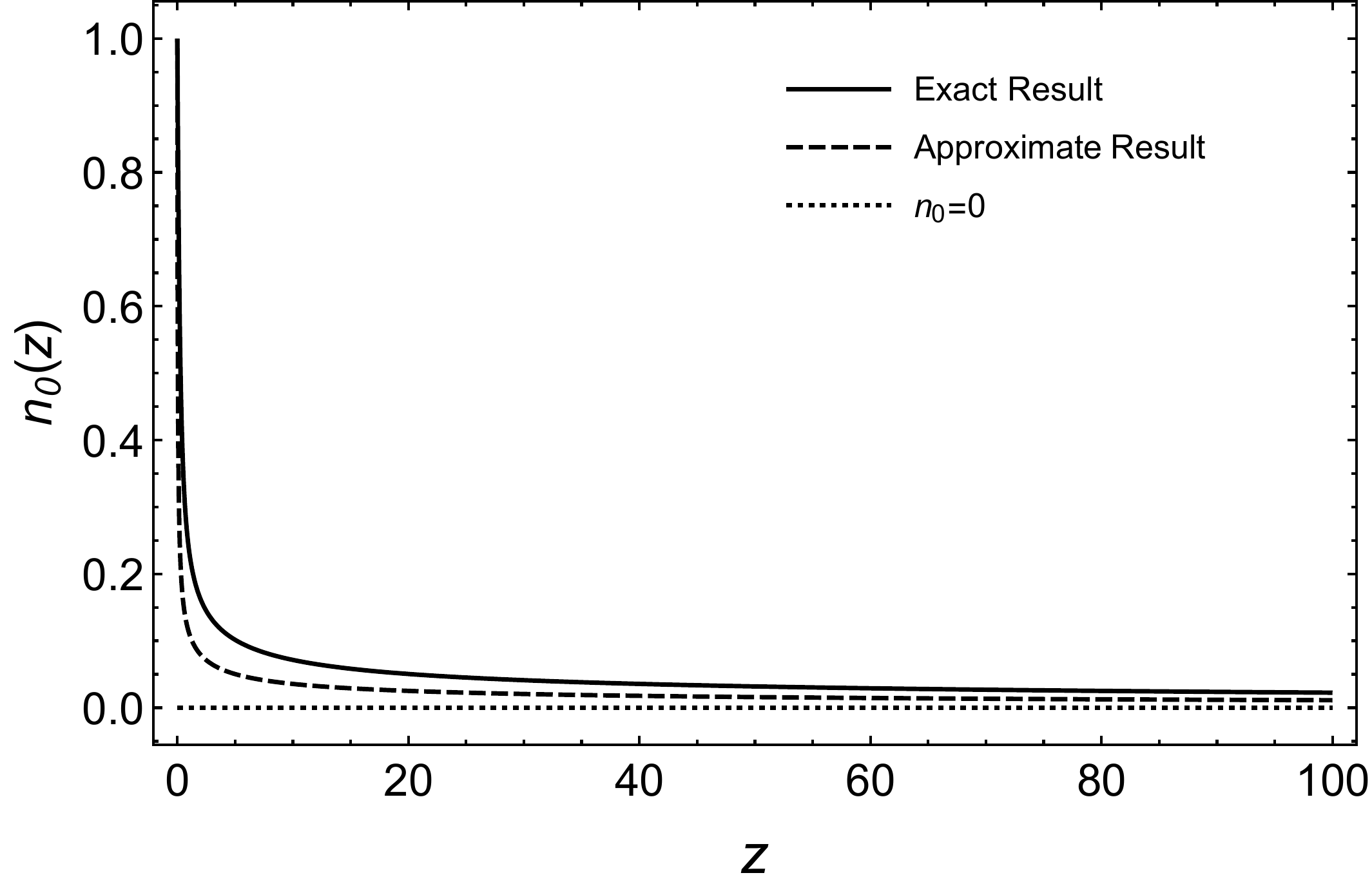} \caption{$n_0(z)$ Vs $z$ in the absence of noise. The approximate result supports KZ mechanism i.e., $n_0 \sim 1/\sqrt{z}$, and matches with the exact result in the large-$z$ regime.} \label{fig:n0}
\end{figure}

Integrating it  over the $k$ modes    one obtains  the exact result for the total defect density $n_0$ in the absence of noise,
\begin{equation}
n_0(z)=\frac{1}{\pi}\, \int^{\pi}_{0} d k \,e^{-2\pi z\,\sin^2{k}}=e^{-\pi\,z}\, I_{0}[\pi\,z],
\label{eq:n0_exact}
\end{equation}
where $I_0(\pi\,z)$ is the modified Bessel function of the first kind.  The limit of large $z$ (or the small sweep speed regime) reveals that the defect density scales as $n_0 \sim 1/\sqrt{z}$, which indeed matches with the  prediction of the KZ scaling behavior (see Fig.~\ref{fig:n0}). 

Interestingly, recent numerical studies have shown that the defect density production exhibits completely different scaling when in addition to the usual linear protocol a small Gaussian noise is present in the control field~\cite{PhysRevLett.117.080402, PhysRevB.95.224303}. In particular, the defect density scales as 
\begin{equation}
n\approx c\, \tau_Q^{-1/2}+ d\, \eta^2 \, \tau_Q,
\label{eq:neta}
\end{equation}
where $c$ and $d$ are  system dependent parameters. The first
term in the above equation Eq.~(\ref{eq:neta}) accounts for the 
usual KZ scaling behavior. The second term is 
proportional to the square of noise amplitude  and represents
 increased defect-density production with the increase of $\tau_Q$ (or is inversely proportional to the sweep speed). This  is converse to that of  the KZ scaling and is termed as  the  AKZ
scaling behavior.

In the next section we will consider linear protocol with transverse  noise and obtain   analytical results  for the defect density.   We will show that for large-$z$ values   AKZ scaling behavior dominates. In the subsequent sections, we will calculate the entropy and correlation functions of a driven quantum XY-spin chain in the presence of noise.

\section{Effect of  Noise on the Defect Generation}
The time dynamics of states spanned by  $|0\rangle$ and $|k,-k\rangle$ is governed by the 
Hamiltonian $\mathcal{H}^{\eta}_k (t)$,
\begin{multline}
\mathcal{H}^{\eta}_k(t)= 2\, \begin{bmatrix}
 h(t)+ J\cos(k)& J\, \gamma_0\sin(k)  \\
J\, \gamma_0\sin(k) & -h(t)-J\cos(k) 
\end{bmatrix}\\
+ 2\,\eta(t)\,\,   \begin{bmatrix}
0&J\, \sin(k)  \\
J\,\sin(k) & 0 \label{eq:NoiH}
\end{bmatrix},
\end{multline}
where the first term in the above Hamiltonian is the usual  
deterministic part. The second term couples the 
homogeneous time-dependent Gaussian noise $\eta(t)$ to  each of the  
$k$-mode within the restricted subspace.
In this subspace, a general state at any given time $t$ for a single realization of the noise $\eta(t)$ can be written as, $|\Psi^{\eta}_k(t)\rangle=u^{\eta}_k (t) |0\rangle+ v^{\eta}_k (t) |k,-k\rangle$, where  $u^{\eta}_k (t)$ and $v^{\eta}_k (t)$ are the time-dependent amplitudes. 
The system starts out  in the perfect PM state 
defined by the initial conditions $u^{\eta}_k(-\infty)=1$ and $v^{\eta}_k (-\infty)=0$.
The time evolution of the general 
state $|\Psi^{\eta}_k(t)\rangle$ for a single noise realization is governed by the stochastic Schr\"{o}dinger equation,
\begin{equation}
i\frac{d}{d t} |\Psi^{\eta}_k(t)\rangle= \mathcal{H}^{\eta}_k(t) |\Psi^{\eta}_k(t)\rangle,
\end{equation} 
where the solution  has to be averaged over all possible realizations (ensemble averaging due to the noise) of the 2-level system corresponding to each $k$-mode~\cite{PhysRevB.67.144303, PhysRevB.87.224301, PhysRevB.96.075419}. 

\begin{figure}
\includegraphics[scale=0.62]{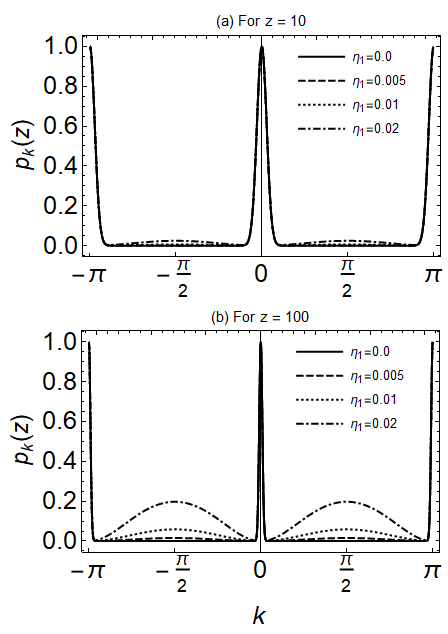} \caption{$p_k$ Vs $k$ in the presence of the fast noise: Interestingly, in the the large-$z$ regime ($z\gg z_1$) or slower sweeps the fast noise begins to affect the system. New critical region (where the excitation probability becomes non-zero) opens up symmetrically around $k=\pm \pi/2$ resulting in more defect generation. } \label{fig:Pk}
\end{figure}

The projected Hamiltonian Eq.~(\ref{eq:NoiH}) is equivalent to the noisy LZ problem with the noise  present only in the  transverse part of the Hamiltonian. We consider the  density matrix, $\hat{\rho}^{\eta}_k (t)=|\Psi^{\eta}_k(t)\rangle \langle\Psi^{\eta}_k(t)|$, and set up the time evolution equation for the population inversion $\rho^{\eta}_k$ (difference between the unexcited and the excited density for a given mode $k$):
\begin{multline}
\frac{d}{d\tau}\rho^{\eta}_k(\tau) = -\frac{1}{2} \int^{\tau}_{-\infty} d\,\tau_1 e^{i\, \int^{\tau}_{\tau_1} d \tau_2\, (v_{\mathrm{LZ}} \tau_2) /2}
\,\rho^{\eta}_k(\tau_1)\\
-\frac{1}{2\,\gamma_0^2}  \int^{\tau}_{-\infty} d\,\tau_1 e^{i\, \int^{\tau}_{\tau_1} d\tau_2\, (v_{\mathrm{LZ}} \tau_2 )/2} \\ 
\times  \eta(\tau) \eta(\tau_1) \,\rho^{\eta}_k(\tau_1) + \text{h.c.},
\end{multline}
where $\tau=2 \Delta_k (t +\cos{k}/v)$, and
 $v_{LZ}=v/ \Delta^{2}_k$.
Taking the noise average one obtains
\begin{eqnarray}
&&\frac{d}{d\tau}{\rho_k(\tau)} = - \int^{\tau}_{-\infty} d\,\tau_1 \cos[\frac{v_{LZ}}{4}(\tau^2-\tau^2_1)]
\rho_k(\tau_1)  \notag{}\\
&&-\frac{1}{\gamma_0^2}\int^{\tau}_{-\infty} d\,\tau_1\cos[\frac{v_{LZ}}{4}(\tau^2-\tau^2_1)] \overline{\eta(\tau) \eta(\tau_1)}\,\rho_k(\tau_1),
\end{eqnarray}
where $\rho_k(\tau_1)$ is obtained by performing  noise average over $ \overline{\rho_k^\eta(\tau_1)}$.  The fast noise criteria allows us to decouple  $\overline{\eta(\tau) \eta(\tau_1)\rho^\eta_k(\tau_1)}$ into a separate product of the noise terms and  the density matrix term, $ \overline{\eta(\tau) \eta(\tau_1)}\,\rho_k(\tau_1)$~\cite{PhysRevB.87.224301}.
The solution of the reduced master equation in the $t\rightarrow  \infty$
limit  is obtained by following the approach of Ref.~\onlinecite{PhysRevB.87.224301} and is given by $\rho_k=  e^{-2\pi\, \eta_0^2/(2\,v_{LZ}\, \gamma_0^2)}(2\,e^{-\pi/(2\, v_{LZ})}-1)$. The noise averaged excitation probability thus obtained is 
\begin{equation}
p_k(z)= \frac{1}{2}\left[1+ e^{-4\pi z\,\eta^2_1 \sin^2{k}}\,(2\,e^{-2\pi z \sin^2{k}}-1) \right],
\label{eq:Pk}
\end{equation}
where  $\eta_1=\eta_0/\gamma_0$.
It is interesting to note that the excitation 
probability [Eq.~(\ref{eq:Pk})] which is non-zero around the $k=0,\pm \pi$ points in the absence of noise, also opens up  around  $k=\pm \pi/2$ regions  in the presence of  noise and in particular for the large $z$-limit or the small sweep speed scenario, which is the result  of dephasing  due  to  the  presence  of  the  fast  noise which is one of the main reason for more defect generation for slower sweeps  (see Fig.~\ref{fig:Pk}).  Recently, similar results have been reported in Ref.~\onlinecite{PhysRevA.103.012608} for the noisy drive protocols in a trapped ion experiment. It is worth noting that the fast sweep regime ($z\ll z_1$) is unaffected by the fast noise.

\begin{figure}
\includegraphics[scale=0.55]{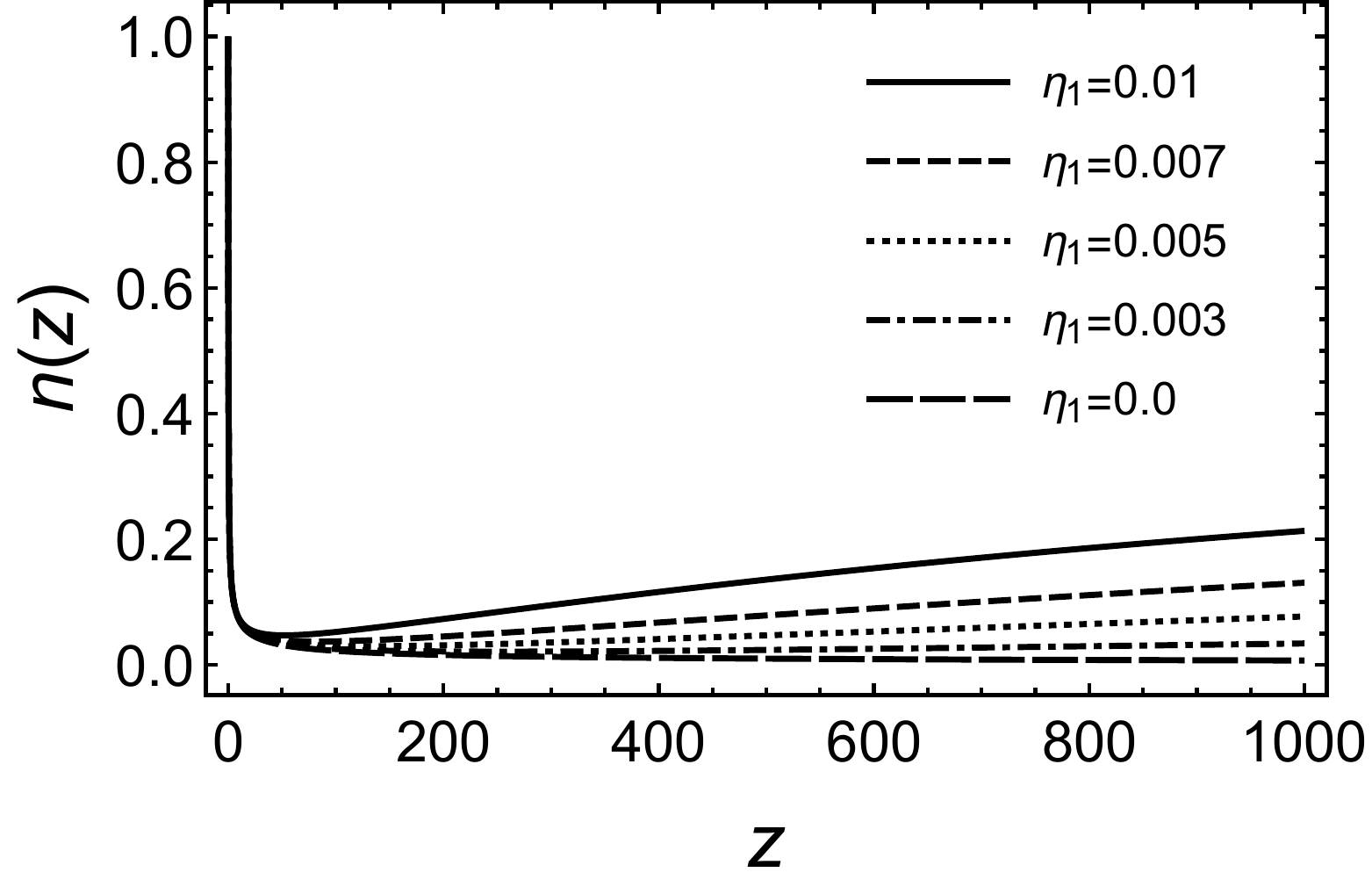} \caption{Effect of the fast noise on the defect density: The defect density has been plotted with respect to $z$ for different noise strength, which is consistent with the AKZ picture for large-$z$ regime i.e., enhanced defect generation for slower sweeps beyond the optimal quench rate which depends on the noise strength. } \label{fig:n}
\end{figure}

\begin{figure}
\includegraphics[scale=0.57]{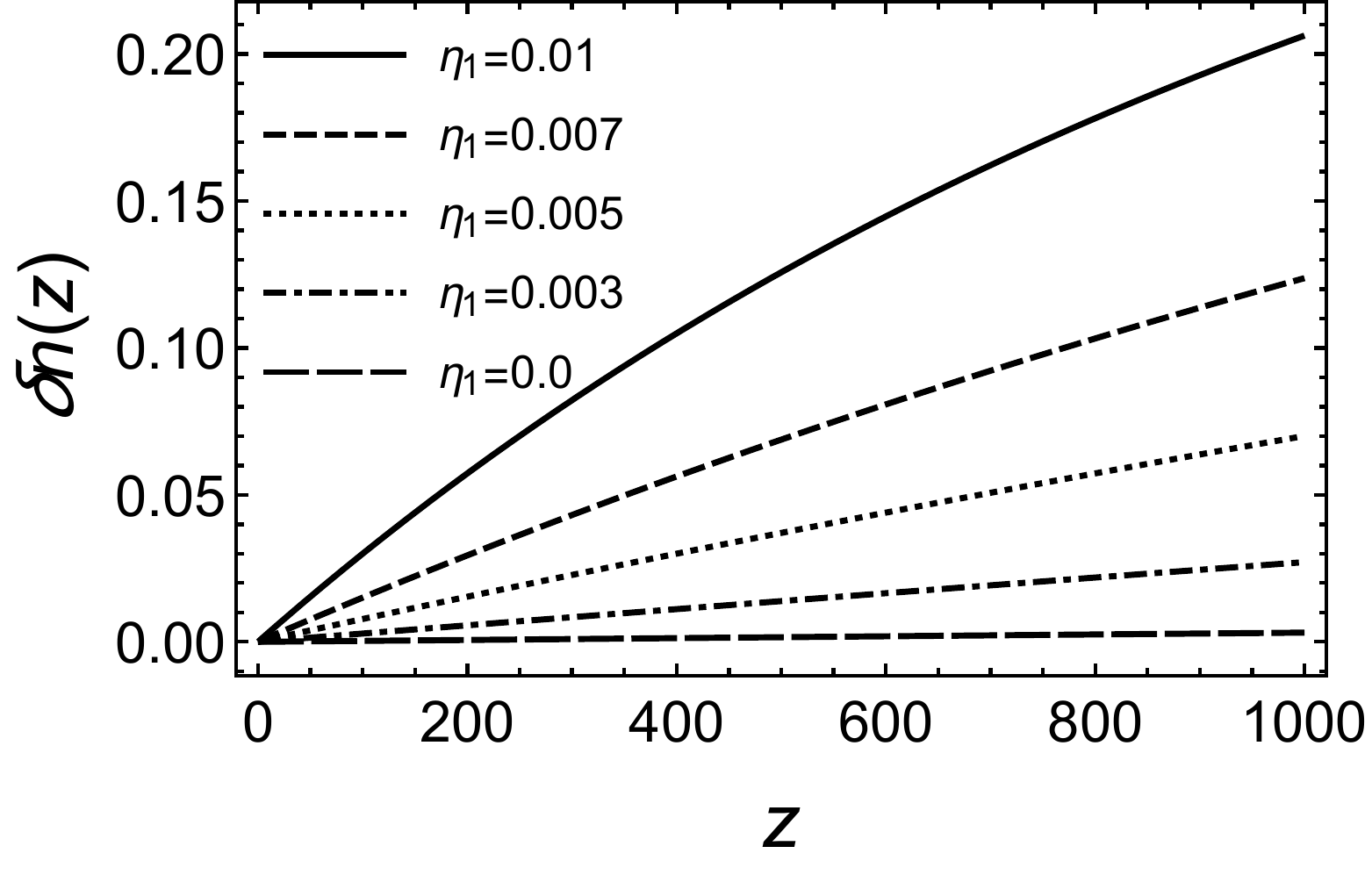} \caption{ Anti-Kibble-Zurek scaling behavior of the defect density: The difference, $\delta n= n(z)-n_0(z)$, scales linearly with $z$ for different noise strengths for $\eta_1 \ll 1$ and for $z \eta^2_1 \sim 1$ (large-$z$) regime.} \label{fig:dn}
\end{figure}

We next evaluate the defect density 
by  integrating $n(z)=\int^{\pi}_{0} d k\, p_k/\pi$
and obtain
\begin{eqnarray}
n(z)=\frac{1}{2} +  e^{-\pi \,(z+\bar{z})}\,I_{0}[\pi \,(z+\bar{z})] -  \frac{1}{2} e^{-\pi \bar{z}} I_{0}[\pi\bar{z}],
\label{eq:n}
\end{eqnarray}
where $\bar{z}=2z\eta_1^2$. 
In the limit, 
$\eta_1\ll  1$, the defect density is approximated as
\begin{equation}
n(z)\approx \frac{1}{\sqrt{2}\,\pi\,\sqrt{z}}+\pi \eta_1^2 z,
\label{eq:n_AKZ}
\end{equation}
which gives the AKZ scaling behavior (see Figs.~\ref{fig:n} and~\ref{fig:dn}). 
From the above expression it can be deduced that the defect density is minimized for a optimal quench rate given by, 
\begin{equation}
z_{\mathrm{O}} \approx 2\sqrt{2}\, \pi^2 \eta^{-4/3}_1 \,\propto \,\eta^{-4/3}_1 .
\label{eq:optimal_t}
\end{equation}
We note that the optimal quench time has a universal universal power-law dependence on the noise strength.


\section{Decoherence of Local Observables}
In both the noiseless and noisy drive  protocols, the XY-spin chain (with $N$ spins) is prepared in a PM state at the initial time $t_{\mathrm{in}}=-T$. This initial state is a pure state i.e., the full system density matrix can be written as $\rho(t=-T)=|0_N\rangle\langle 0_N|$. Subsequently the system is driven by the transverse magnetic field $h(t)=t/\tau_Q$ through the quantum critical points ($h=\mp J$) up to the final time $t_{f}=T$. In the noiseless drive scenario the full density matrix of the evolved $N$-spin chain remains in the pure state due to the unitary time evolution. However, for large system size ($N\rightarrow \infty$) and in the long time limit i.e., $T\rightarrow \infty$, the coherences of the density matrix develop highly fluctuating phases (dependent on $k$ and $T$) which vanishes when integrated over $k$ for all local observables~\cite{PhysRevA.73.043614}. 
 This decohered density matrix corresponds to the nonequilibrium steady state (NESS) which is fundamentally different from the decoherence process due to any external or internal noise~\cite{PhysRevA.73.043614}.
In addition to the internal decoherence the dephasing is further enhanced by the noise in the drive protocol which results in the exponential suppression of the fluctuating coherences. 

For $\eta^2_0 J \ll \Delta_k$  the noise has negligible effect on the system. The crossover region is around  $\eta^2_0 J \sim \Delta_k$ at which the noise begins to play a role in the dynamics of the system. In the limit $\eta^2_0 J \gg \Delta_k$, the fast noise effects the system the most specifically in the non-adiabatic regions (when the gap $\Delta_k \rightarrow 0$) around the quantum critical points ($k=0,\pm \pi$), in addition new critical regions around $k=\pm \pi/2$ become important.
Overall, non-adiabatic effects are enhanced due to the noise which gives incoherent contributions to the defect density leading to increased defect generation at the end of the protocol.

In the long time limit, the noise averaged off-diagonal terms of the density matrix (coherences), 
$\bar{\rho}^k_{12}$ and $\bar{\rho}^k_{21}$ vanish.
Therefore the noise averaged  decohered density matrix, $\bar{\rho}_D=\otimes_{k>0}\bar{\rho}_{D,k}$ (with $\bar{\rho}_{D,k}$ a diagonal $2\times 2$ matrix in the subspace $|0\rangle,|k,-k\rangle$) in the limit $N\rightarrow \infty$ and $T\rightarrow \infty$, with respect to the final decohered state 
can be written as, 
\begin{equation}
\bar{\rho}_{D,k}=\begin{pmatrix}
p_k & 0\\
0 & 1-p_k
\end{pmatrix}.\label{eq:rho_decohered}
\end{equation}
We will use the above expression of the density matrix to calculate the noise averaged observables in the final decohered state.

\subsection*{ 2-Point Correlator in the fermionic representation}
Similar to Ref.~\onlinecite{PhysRevA.73.043614}, one can define a $2$-point correlator in  terms of the  fermionic operators  as, 
 \begin{equation}
d\,(x-x')= \langle c_x\, c^{\dagger}_{x'}  \rangle= \frac{1}{2\pi}\int^{\pi}_{-\pi} d k\, e^{-i\,k\,(x-x')}  \,p_k .
 \end{equation}
 This correlator is non-zero for $x-x'=2n$, where $n$ is an integer. In the absence of  noise  the 2-point correlator is
\begin{equation}
d_n(z)= \frac{1}{2\pi}\int^{\pi}_{-\pi} d k\, e^{-i\,k\,n}  \,p^{0}_k =e^{-\pi\,z}\,I_n(\pi\,z),
\end{equation}
where $I_n(\pi\,z)$ is the modified Bessel function of the first kind. The large-$n$ expansion of $d_n$  at fixed $z$ yields
\begin{eqnarray}
 d_n(z) \approx \frac{1}{2\pi}\int^{\pi}_{-\pi}dk\, e^{-\pi\,z\, k^2/2} e^{i\,n\,k}
 =\frac{e^{-n^2/2\pi z}}{\sqrt{2 \pi^2 z}}.
\end{eqnarray}
Thus consistent with the KZ-picture the correlation length of  the density correlator  is proportional to $\sqrt{z}$ and the magnitude of the correlator is inversely proportional to $\sqrt{z}$.

A similar calculation for the 2-point correlator in the presence of noise, $p_k$ yields  
\begin{multline}
d_n(z)= \frac{1}{2\pi}\int^{\pi}_{-\pi} d k\, e^{-i\,k\,n}  \,p_k \\
=\frac{1}{2}\delta_{n,0}-\frac{1}{2}  e^{-\pi \bar{z}} I_{n}[\pi  \bar{z}] + e^{-\pi (z +  \bar{z}) }\,I_{n}[\pi (z +  \bar{z}) ].
\label{eq:den_corr_1}
\end{multline}
In the limit of $\eta_1\ll 1$ and $\pi \bar{z}< 1$  and for  large-$n$ expansion ($n> \sqrt{4\pi z\eta_1^2 \ln1/\eta_1}$) the above expression  is approximated as, 
\begin{equation}
d_n(z)\approx \frac{1}{\sqrt{2\pi^2 (z+\bar{z})}} e^{-n^2/[2\pi (z+\bar{z})]}.
\label{eq:den_corr_3}
\end{equation}
 Thus the  correlation length in the presence of noise $l_{n}=\sqrt{\pi (z+\bar{z})}$ remains proportional to $\sqrt{z}$ and 
increases with the strength of the noise.

\section{Entropy Density in The Final Decohered State}
\begin{figure}
\includegraphics[scale=0.5]{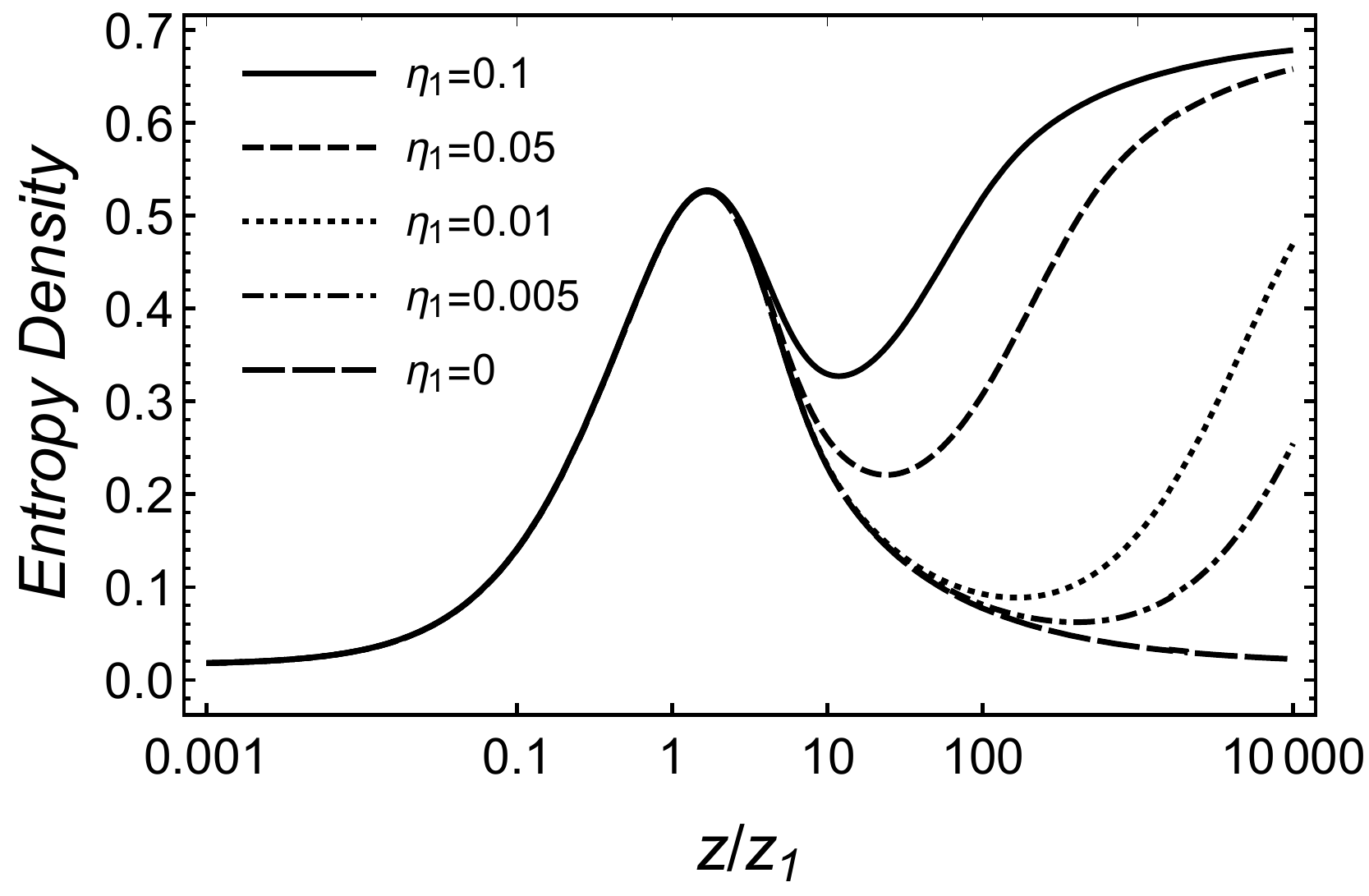} \caption{Entropy density Vs. $z$ plotted for different noise strengths. The entropy density is consistent with the AKZ picture. The entropy density increases after the optimal quench time which is the signature of increased defect generation due to the fast noise in slower sweep regime. For very slow sweeps, the noise (with $\eta_1 \ll 1$) can randomize the system to the maximally mixed state (i.e. $S/N \rightarrow \text{ln}\,2$). Apart from that, the entropy density maximizes (locally) at $z=z_1$ which is the signature of the crossover behavior of the spin correlation functions from monotonically decreasing behavior for $z<z_1$ to the oscillatory behavior for $z>z_1$ as discussed in Ref.~\onlinecite{PhysRevA.73.043614}. } \label{fig:entropy}
\end{figure}

The decohered state has a finite entropy density which is a clear indication that the final state is a mixed state. To quantify the amount of information lost in the decoherence process (at the end of the drive protocol) we calculate the Von-Neumann entropy ($S=-N \text{tr}\,\rho_D \ln \rho_D$) in terms of the excitation probability as follow, 
\begin{equation}
 S=-\frac{N}{2\pi}\int^{\pi}_{-\pi} d k [p_k \ln p_k  +(1-p_k ) \ln(1-p_k )],
\end{equation}
where $p_k$ is given in  Eq~[\ref{eq:Pk}]. The above integration is performed by expanding both the log terms in terms of $e^{-4\pi z \eta^2_1 \sin^2{k}}(2e^{-2\pi z \sin^2{k}}-1)$ and integrating each of the terms individually. The final result of the entropy density ($S/N$) can be expressed in the following series form, 
\begin{equation}
S/N=\ln 2 - \sum^{\infty}_{m=1}\sum^{2 m}_{r=0} \frac{2^{2m-r}(-1)^r \binom{2m}{r}}{2m\,(2m-1)} e^{-\pi z Y^r_m}
 I_{0}[\pi z Y^r_m],
 \label{eq:entropy}
\end{equation}
where $Y^r_m=(2m-r + 4m\eta^2_1 )$. 
In Fig.~\ref{fig:entropy} we plot Eq.~(\ref{eq:entropy}) for different noise strengths. One can observe from the figure that the finite entropy density depends on the sweep speed and also that for each noise strength their exists an optimal quench rate either side of which the entropy density increases. In particular for $z>z_{\mathrm{O}}$  (where $z_{\mathrm{O}}$ is the optimal quench rate) the entropy increases which is the signature of AKZ behavior of defect production. 
 For $\eta_1\neq 0$, the entropy density asymptotically approaches $\ln 2$ for large-$z$ value i.e., a fully mixed state is formed or in other words, the system approaches an asymptotic infinite temperature steady state.
 However, for the fast sweep speeds ($z<z_1$), the driven system is not affected by the noise. In the intermediate region the system has some finite non-zero entropy density which signifies a partially mixed state.
 
\section{Magnetization and Spin-spin Correlations}
The expectation value of the spin-spin correlators with respect to the decohered state is conveniently obtained in terms of the pair products of Majorana fermion operators. 
Consider the Majorana fermion operators~\cite{LIEB1961407, PhysRevA.3.786, PhysRevA.73.043614, PhysRevB.89.024303}
\begin{equation}
A_x=c^{\dagger}_x+c_x,\,\,B_x=c^{\dagger}_x-c_x.
\end{equation}
The pair product of spins $\sigma^{\alpha}_x\,\sigma^{\alpha}_{x+n}$ and that of the Jordan-Wigner string variable $\tau_x \tau_{x+n}$ (with $\tau_x=\Pi_{x<x'}(-\sigma^3_{x'})$) can be expressed as follows~\cite{ PhysRevA.73.043614},
\begin{align}
\sigma^{1}_x\,\sigma^{1}_{x+n}=B_x A_{x+1} B_{x+1}....A_{x+n-1}B_{x+n-1}A_{x+n},\\
\sigma^{2}_x\,\sigma^{2}_{x+n}=A_x A_{x+1} B_{x+1}....A_{x+n-1}B_{x+n-1}B_{x+n},\\
\tau_{x}\tau_{x+n}=A_x B_x A_{x+1}B_{x+1}....A_{x+n}B_{x+n}.
\label{eq:full_latt_corr}
\end{align}
Due to the sub-lattice structure of the decohered matrix i.e. decoupling of the decohered matrix, $\rho_D=\rho_E \otimes \rho_O$, with $\rho_E$ and $\rho_O$ acting correspondingly on the even and odd sub-lattice only, the expectation values of the full lattice spin correlators can be written as a product of sublattice correlators as follow~\cite{ PhysRevA.73.043614},
\begin{equation}
\langle\sigma^{\alpha}_x\,\sigma^{\alpha}_{x+2n}\rangle =\langle\langle\sigma^{\alpha}_x\,\sigma^{\alpha}_{x+n}\rangle\rangle \langle\langle\tau_x\,\tau_{x+n}\rangle\rangle,\,\,\alpha=1,2.
\label{eq:full_corr}
\end{equation}
\begin{equation}
\langle\tau_x\,\tau_{x+2n}\rangle = \langle\langle\tau_x\,\tau_{x+n}\rangle\rangle\,\langle\langle\tau_x\,\tau_{x+n}\rangle\rangle,
\end{equation}
where $\langle...\rangle$ and $\langle\langle...\rangle\rangle$ represent the expectation values  on the full lattice and the sublattice, respectively. The sublattice correlators are obtained using the Wick's theorem which requires one to evaluate correlators of the form $\langle B_x A_{x'} \rangle$, $\langle A_x B_{x'}\rangle$, $\langle A_x A_{x'}\rangle$ and $\langle B_x B_{x'} \rangle$. Out of these only the correlators of the type $\langle B_x A_{x'} \rangle$, $\langle A_x B_{x'}\rangle $ are needed to calculate the sublattice correlators which can be expressed in the form of the Toeplitz determinants.

The expectation value of the magnetization operator, $m_z$, evaluated with respect to the final decohered state and in terms of the Majorana operators is given by,
\begin{equation}
m_z=\langle \sigma^3_x\rangle=\langle A_x B_x\rangle,
\end{equation}
and the expectation value of the magnetization correlator is given by,
\begin{equation}
\langle \sigma^3_x \,\sigma^3_{x'} \rangle=\langle A_x B_x A_{x'} B_{x'} \rangle.
\end{equation}
We first evaluate the magnetization and magnetization correlators following which we discuss $\sigma^{1,2}_x$ spin-spin correlators.
\subsection{Magnetization Density and Magnetization Correlation}
The average magnetization density $\langle\sigma^3_x\rangle$ at the end of the quench protocol is given by,
\begin{equation}
m^z=\langle A_x\,B_x \rangle=1-2 n(z),
\label{eq:mz}
\end{equation}
where $n(z)$ is the noise dependent defect density [Eq.~(\ref{eq:n})] and in terms of which,
\begin{equation}
m^z=  e^{-\pi \bar{z}} I_{0}[\pi\bar{z}]
 -2\, e^{-\pi(z+\bar{z})}\,I_{0}[\pi (z+\bar{z})].
\label{eq:mz_noisy}
\end{equation}
\begin{figure}
\includegraphics[scale=0.52]{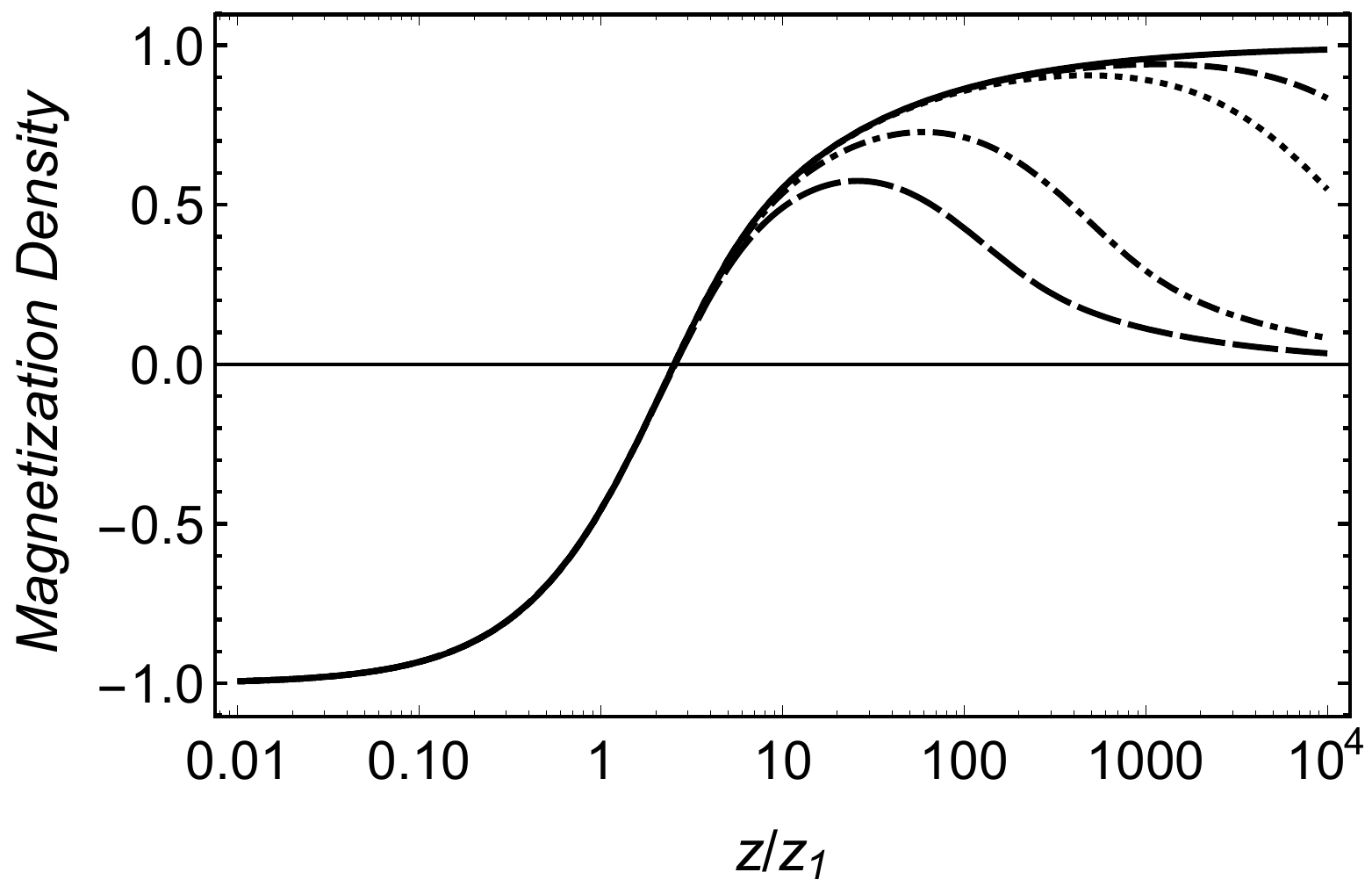} \caption{Magnetization density as a function of $z$ reduces after the optimal quench time/rate in agreement with the AKZ picture. For very large-$z$ case the noise can randomize the system completely resulting in a zero magnetization density.} \label{fig:mz}
\end{figure}
Thus the large-$z$ limit of the magnetization density given by,
\begin{equation}
m_z-1 \approx -\frac{1}{\sqrt{2}\,\pi\,\sqrt{z}}-\pi z \eta^2_1,
\end{equation}
 is consistent with the AKZ picture or in other words, it decreases after the optimal quench time, Eq~[\ref{eq:optimal_t}], when the defect production starts to increase due to the noise (see Fig.~\ref{fig:mz}).

Consider next the magnetization correlator in the $z$-direction, $\langle \sigma^3_x \, \sigma^3_{x+2n} \rangle$, given by,
\begin{equation}
\langle \sigma^3_x \, \sigma^3_{x+2n} \rangle=\langle \sigma^3_x \rangle^2 - \left(2  \int^{\pi}_{-\pi}\,\frac{dk}{2\pi}\, e^{-ikn} p_k\right)^2.
\end{equation}
The connected correlator $C_n(z)$ is obtained by subtracting the position independent part, $\langle \sigma^3_x \rangle^2$ from $\langle \sigma^3_x \, \sigma^3_{x+2n} \rangle$ and  is given by  $C_n(z)=-4 d^2_n (z) $, where $d_n (z)$ is given by the equation Eq.~(\ref{eq:den_corr_1}). 
Therefore, the magnetization correlation for large-$n$ retains the KZ scaling relation with the 
 correlation length given by 
 \begin{equation}
l_{noisy}=\sqrt{\pi( z +\bar{z})/2},
\end{equation}
where we note that the 
 magnetization correlation length increases 
 as compared to the noiseless scenario. 
 It is interesting to note that the presence of  noise in the anisotropy  decreases the magnetization density due to increased defect production, 
  the correlation length of the magnetization correlator, however,  increases with the strength of the noise.
 The amplitude of the magnetization correlator nevertheless decreases
with the noise.
\subsection{Spin correlators: $\langle\sigma^{1,2}_x\,\sigma^{1,2}_{x+2n}\rangle$}
As shown in  Eq.~(\ref{eq:full_corr}) the spin correlators can be expressed as product of sublattice correlators. One can represent the sublattice correlators at $n$-separation in terms of the determinants of the Toeplitz matrices:
\begin{equation}
\langle\langle\sigma^{1}_x\,\sigma^{1}_{x+n}\rangle\rangle=D_n[g^{+1,z}],
\end{equation}
\begin{equation}
\langle\langle\sigma^{2}_x\,\sigma^{2}_{x+n}\rangle\rangle=D_n[g^{-1,z}],
\end{equation}
\begin{equation}
\langle\langle\tau_x\,\tau_{x+n}\rangle\rangle=D_n[g^{0,z}],
\end{equation}
where $g^{m,z}$ are the generating functions defined as, 
\begin{equation}
g^{m,z}(\xi)=-(-\xi)^m\,(1-2\,p_k),
\label{eq:gen_func0}
\end{equation}
where $\xi=e^{2ik}$
and $D_n[g^{m,z}]$ are the corresponding Toeplitz matrix determinants for different sublattice spin correlators. Given the generating function $g^{m,z}$ the determinant $D_n[g^{m,z}]$ are defined as~\cite{PhysRevA.73.043614, PhysRevB.89.024303},
\begin{equation}
D_n[g^{m,z}]=\begin{vmatrix}
f^{(m)}_0 & f^{(m)}_{-1} &....& f^{(m)}_{-(n-1)}\\
f^{(m)}_1 & f^{(m)}_{0} &....& f^{(m)}_{-(n-2)}\\
. & . &....& .\\
. & . &....& .\\
. & . &....& .\\
f^{(m)}_{n-1} & f^{(m)}_{n-2} &....& f^{(m)}_{0},
\end{vmatrix}
\label{eq:T_determinant}
\end{equation}
where the elements of the determinant $f^{(m)}_l$ is the $l^{\text{th}}$ cummulant of the generating function $g^{m,z}(\xi)=\sum_l f^{(m)}_l \, \xi^l $ and are obtained by performing the following contour integration,
\begin{equation} 
 f^{(m)}_l=\oint_{C} \, \frac{d \xi}{2 \pi i \xi}\,\xi^{-l}\,g^{m,z}(\xi), 
 \label{eq:gen_func1}
\end{equation}
where $C$ is a unit circle contour with $|\xi|=1$. The above integral in terms of the $k$-variable acquires the form,
\begin{equation}
f^{(m)}_l= \int^{\pi/2}_{-\pi/2}\, \frac{dk}{\pi}\,e^{-i2k l}\,g^{m,z}(e^{i2k}).
\label{eq:gen_z_T_elements}
\end{equation}
The integral is evaluated by taking  the integral representation of the modified Bessel function of the first kind,
\begin{equation}
I_\nu(z)=\frac{1}{2\pi}\int_{-\pi}^{\pi}d\theta e^{z\cos\theta-i\nu \theta}
\end{equation}
where $\nu$ is an integer and $\text{Re}(z)>0$. Using the above
 integral identity yields
 \begin{align}
f^{(m)}_{l}=\frac{e^{-\pi \bar{z}}}{2 \pi} \,I_{l-m}[\pi\bar{z}]
-\frac{\,e^{-\pi (z+\bar{z})}}{\pi}\,I_{l-m}[\pi (z+\bar{z})],\label{eq:noisy_Telements} 
\end{align}
with $m=0,\pm 1$. In the large-$z$ limit (and with $z\eta^2 \neq 0$), the expression reduces to 
\begin{equation}
f^{(m)}_{l}= (-1)^m\frac{1}{\pi}\left[-\frac{e^{-(l-m)^2/2\pi \bar{z}}}{\sqrt{2\bar{z}}}
+\frac{\sqrt{2}\,e^{-(l-m)^2/2\pi  (z+\bar{z})}}{\sqrt{z+\bar{z}}}\right].
\end{equation}
\begin{figure}
\includegraphics[scale=0.18]{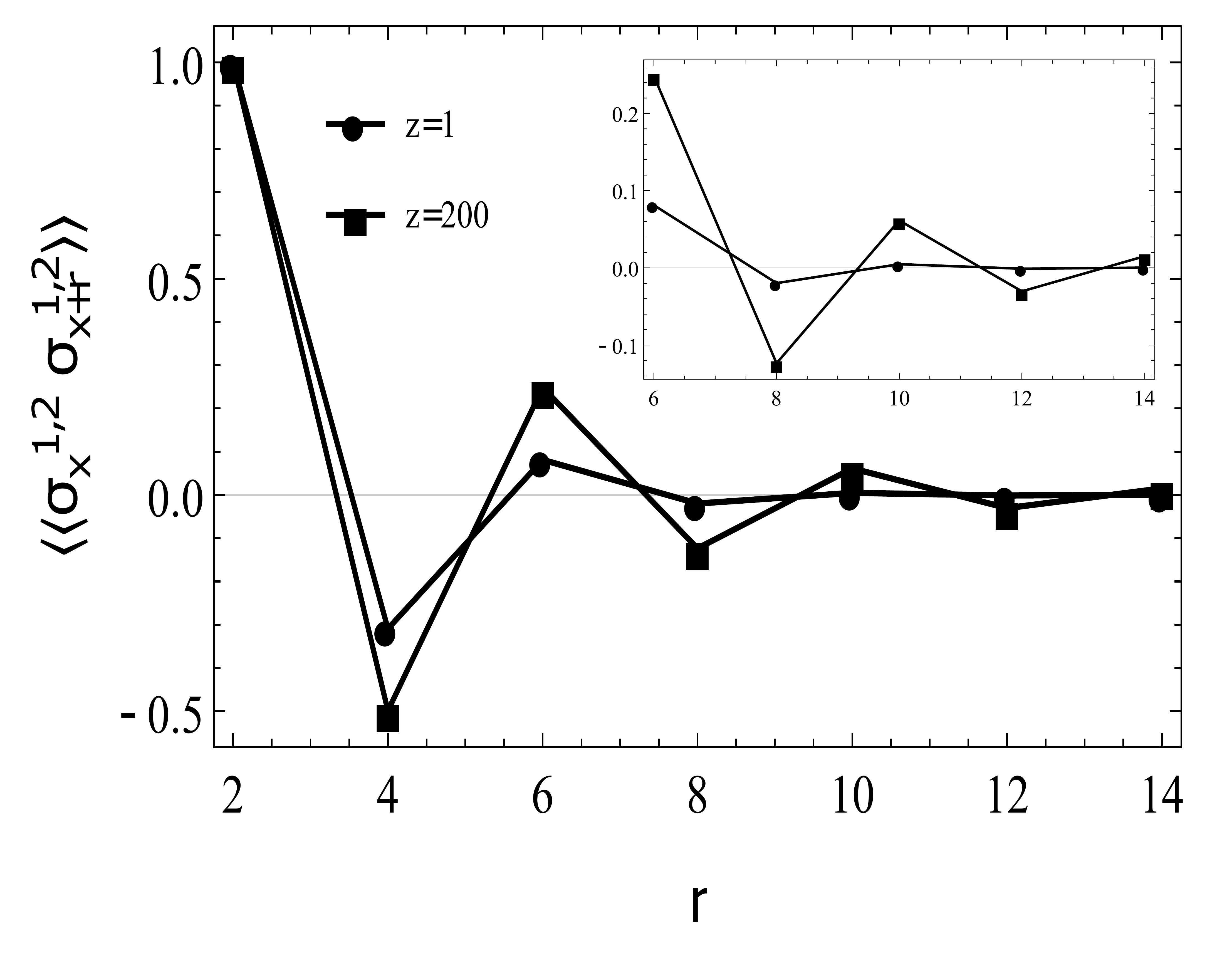} \caption{Sub-lattice spin correlators $\langle\langle \sigma^{1,2}_{x} \sigma^{1,2}_{x+r} \rangle\rangle$  have been plotted with respect to the separation-$r$ 
in the noiseless scenario. The correlators in the figure are normalized from their 
respective value at separation $r=2$.
From the figure one can observe that the larger $z$-values correspond to the large correlation length which is consistent with the KZ behavior. The correlation length ($l_{\sigma}$) becomes small (short ranged spin correlations) in the fast sweep regime. From the inset  one can  notice the  KZ behavior at relatively larger separation. } \label{fig:sigma_corr}
\end{figure}

 \begin{figure}
\includegraphics[scale=0.18]{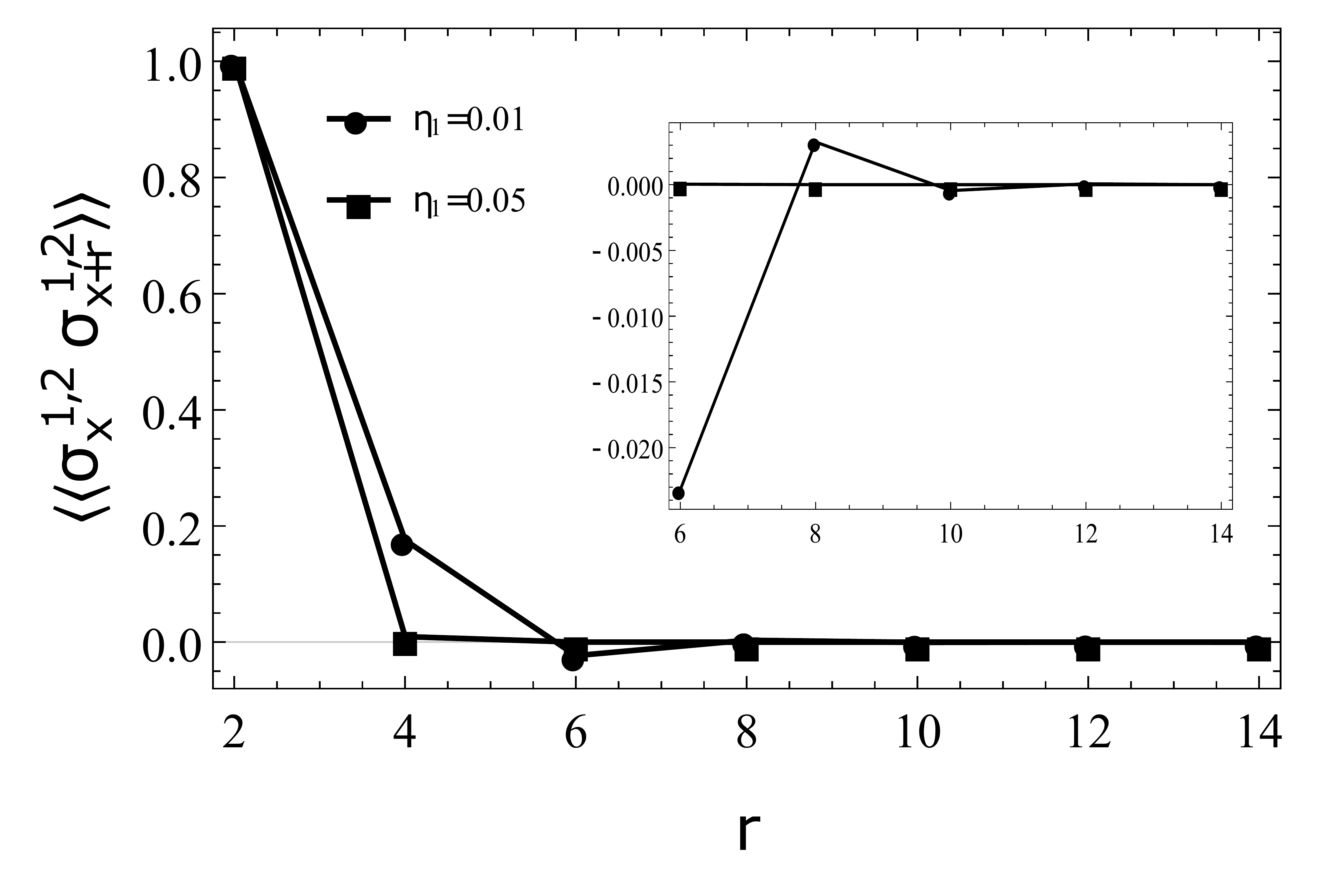} \caption{Normalized $\langle\langle \sigma^{1,2}_{x} \sigma^{1,2}_{x+r} \rangle\rangle$ Vs. $r$ for  different noise strengths ($\eta_1= 0.01\,\, \text{and}\,\, 0.05$) at $z=200$ (slow sweep regime).  The correlation length $l_{\sigma}$  decreases with the increased noise strength, which is the signature of AKZ scaling behavior. } \label{fig:sigma_corr_noisy}
\end{figure}
 In the Figs.~\ref{fig:sigma_corr} and~\ref{fig:sigma_corr_noisy} we plot the numerically calculated determinants for the sub-lattice correlators $\langle\langle\sigma^{1,2}_x\,\sigma^{1,2}_{x+n}\rangle\rangle$. Figure ~\ref{fig:sigma_corr} corresponds to the noiseless case where we observe the KZ scaling behavior for the correlation length, i.e., larger correlation length for smaller sweep speeds. In Fig.~\ref{fig:sigma_corr_noisy} we consider fast noise and use the elements of the Toeplitz matrix  given by Eq.~(\ref{eq:noisy_Telements}) to evaluate the determinants. Here we notice the signature of the AKZ behavior, i.e.,  decrease in the correlation length with the increasing strength of the noise for slower sweeps. Similar behavior is found for the $\langle\langle\tau_x\,\tau_{x+n}\rangle\rangle$ correlator.

\subsection{Spin Correlations at Large Separation}
In this section, we investigate the behavior of spin correlators at large separation in the presence of the fast noise. In the asymptotic limit, following Szeg\"{o}'s limit theorem, the Toeplitz determinant acquires the form~\cite{PhysRevA.73.043614, PhysRevB.89.024303, BASOR1994129, doi:10.1063/1.1699484,PhysRevA.3.786, LIEB1961407},
\begin{equation}
D_n[g^{m,z} (\xi)]\approx exp\left[n\int^{ \pi}_0 \,\frac{d\theta}{ \pi}\, \mathrm{ln} g^{m,z}(e^{i 2\theta})  \right],
\label{eq:Szego}
\end{equation}
where $g^{m,z}(\xi)$ is the generating function  given in Eq.~(\ref{eq:gen_func0}).

The zeroes of the generating function 
plays an important role in the analyticity of the asymptotic behavior of the spin correlators and it can be easily verified that it  has the same set of zeroes as  the  noiseless generating function. 
Therefore, the effect of zeroes of the generating function on  the analyticity of the spin correlators with respect to $z$ will remain the same as for the noiseless drive case. The difference is that the generating function with noisy drive is multiplied by an extra noise dependent exponential factor, $e^{-2\pi\,z\,\eta^2_1(1-x)}$.
We are mainly interested in the role of this extra term on the correlation lengths of the sublattice and the full lattice spin correlators. In the following discussion    we will use the method developed by Cherng and Levitov (\onlinecite{PhysRevA.73.043614})
to show that this extra term is responsible for the AKZ scaling behavior  of the correlation lengths.

The generating function  for the noisy drive can be written as
\begin{equation}
g^{m,z}_{\mathrm{fn}}(\xi)=-(-\xi)^{m}\,\lambda^{-1}_0\,(1-\lambda_0\,\xi)(1-\lambda_0\,\xi^{-1})\,e^{H(\xi)},
\end{equation}
where
\begin{equation}
H(\xi)=h(\xi)-2\pi\,z\,\eta^2_1(1-x),
\label{eq:H_x}
\end{equation}
and 
\begin{equation}
h(\xi)=\mathrm{ln}\,\left[\frac{1-e^{-\pi\,z(1-2\,z_1/z -x)}}{2(1-2\,z_1/z -x)} \right].
\end{equation}
The zeroes closest to  to the   unit circle  are denoted by $\lambda_0$ and $\lambda_0^{-1}$, where
 $\lambda_0 =1/\mathrm{exp}\left[\mathrm{cosh}^{-1}\left( 1-2 z_1/z \right) \right] $. 
Both $h({\xi})$ and $H({\xi})$ can be expanded as a function of $x=(\xi+\xi^{-1})/2$ as follows,
\begin{equation}
h(x)=\sum_{n \geq 0}\,h_n\,x^n\, ,\,
H(x)=\sum_{n \geq 0}\,H_n\,x^n,
\end{equation}
where we notice that $H_0=h_0 -2\pi\,z\,\eta^2_1 $ and $H_{1}=h_1+2\pi\,z\,\eta^2_1$, while for $n\neq 0, 1$,  all other coefficients satisfy $H_n=h_n$.
The  correlation length of the sub-lattice spin correlators are given by the following  expression
\begin{align}
-l^{-1}_{\sigma/\tau}= \int^{1}_{-1}\, \frac{dx}{\sqrt{1-x^2}}\,H(x)= \int^{1}_{-1}\, \frac{dx}{\sqrt{1-x^2}}\,h(x)+\pi^2 \bar{z},
\label{eq:corr_lengths_sub_fn}
\end{align}
where the full lattice correlator is given by 
$l^{-1} = l^{-1}_{\sigma}+l^{-1}_{\tau}$.
\begin{figure}
\includegraphics[scale=0.85]{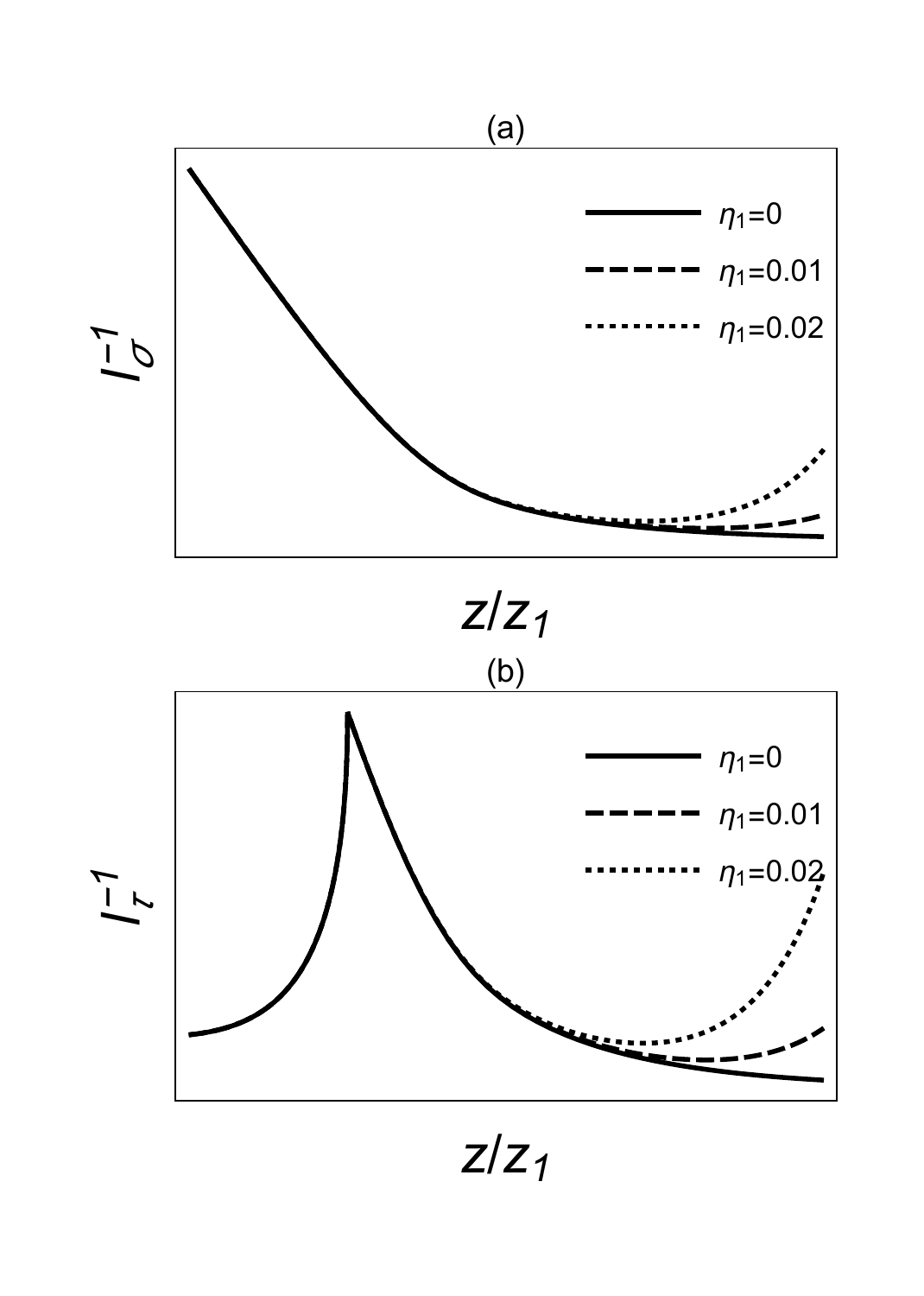} \caption{The sub-lattice inverse correlation lengths have been plotted as a function of $z$ for the different noise strengths, both the $l^{-1}_{\sigma}$ and $l^{-1}_{\tau}$ show the anti-Kibble-Zurek scaling behavior i.e.,  the sub-lattice correlation lengths $l_{\sigma,\tau}$  decreases after the optimal quench rate which depends on the noise strength. Furthermore, the $l^{-1}_{\tau}$ shows non-analytical behavior at $z=z_1$ and beyond $z>z_1$ both   have same values.  Therefore,  it is understood  that the increased noise strength of the fast noise further decreases $l$, making the spin correlators relatively short ranged in the presence of the fast noise. The $l^{-1}=l^{-1}_{\sigma}+l^{-1}_{\tau}$ also shows the non-analytical behavior at $z=z_1$ as a result of the non-analytical behavior of $l^{-1}_{\tau}$, this is the indication of the crossover of different behaviors of the spin correlators i.e. non-oscillatory monotonically decreasing behavior for $z<z_1$ regime and exponentially suppressed oscillatory behavior of spin correlators for $z> z_1$ regime  with respect to the spatial separation.} \label{fig:sub_corr_lengths}
\end{figure}
The result of the above integration yields
\begin{equation}
l^{-1}_{\sigma/\tau}\approx  \sum^{\infty}_{m=0}\,\frac{\Gamma(m+1/2)^2\,\mathrm{Re}\,Li_{m+3/2}(2)}{\pi^{3/2}\,\Gamma(m+1)(2\pi\,z)^{m+1/2}}+\pi^2\bar{z},
\label{eq:l_sublattice}
\end{equation}
where the first term is due to the noiseless case, Ref.~(\onlinecite{PhysRevA.73.043614}), which in the limit of slow sweep speed results in  the correlation length or the domain size  proportional to $\sqrt{z}$ thus  satisfying the KZ scenario.
 On the other hand, for the  fast noise and large $z$ scenario the scaling of the correlation length is consistent with the AKZ picture suggesting that the domain size  reduces, this result quantifies how the fast noise randomizes the driven system spatially at the end of the protocol. The sub-lattice correlation lengths have been plotted in  Fig.~\ref{fig:sub_corr_lengths} with respect to $z$ for different noise strengths. They clearly show the AKZ behavior in the large-$z$ case,  i.e., beyond the optimal quench rate $z_{\mathrm{O}}$  the inverse correlation length starts to increase. For small-$z$ values the fast noise has negligible effect. 

\section{Concluding Remarks}
In this work  we quantify analytically the effects of the fast Gaussian noise in the driven quantum XY-spin chain. We have considered transverse protocol in which the external transverse magnetic field drives the system linearly with respect to time,  starting from the  paramagnetic phase  to a region with the ferromagnetic phase and finally back to the paramagnetic phase in the presence of the time dependent Gaussian noise in the anisotropy term.
 In this protocol, the system passes through two quantum critical points at $h=\mp J$  where the energy gap vanishes resulting in non-adiabatic effects. We map the problem 
to the noisy LZ problem and in terms of the  density matrix formalism we obtain a  reduced master equation for the population inversion. 
The solution of the equation  in the $T\rightarrow\infty$ limit has been utilized to obtain the  final excitation probability.
  
  The implications for the 
correlators due to the non-equilibrium dynamics of the noisy transverse drive protocol are as follows: first, the fast fluctuating coherences vanishes in the large time limit (even without the noise) due to the internal decoherence arising from the  large system size ($N\rightarrow \infty$) and this allows the course graining in momentum space and transforms the pure state into an entropic state with finite non-zero entropy. The time dependent Gaussian fast noise further exponentially suppresses the highly fluctuating coherences  and in particular affects the  system most when the system passes through the quantum critical points. Finally,  the noise can heat up the population  to the asymptotic infinite temperature state for the slower sweeps or large-$z$ case, maximizing the entropy density to $\log 2$ 
and at the same time minimizing the average magnetization density to zero. The  effects of the noise are minimized when driven at an optimal sweep rate which turns out to scale universally with the strength of the fast noise.

We have analyzed the spin-spin correlation functions (in the presence of noise) at the end of the protocol 
for large separation using the Toeplitz determinant asymptotes at large-$n$. For slow sweep speeds the effect of the fast Gaussian noise on the correlation lengths of the spin correlators at large-separation
reveals behavior 
consistent with the anti-Kibble-Zurek picture. The sub-lattice correlation lengths for $\sigma^{1,2}$ and $\tau$-spin correlators decreases with the strength of the noise according to the anti-Kibble-Zurek scaling behavior. We have also analysed the effect of the fast noise on the magnetization $\sigma^3$-spin correlator. For large $n$-separation we find that the correlation length of the magnetization correlator increases with the strength of noise when $ \eta_1\ll 1$ and $\pi \bar{z} < 1$.

\section{ Acknowledgement}
 S G is grateful to  the Science and Engineering Research Board, Government of India, for the support
via the Core Research Grant  Number CRG/2020/002731.

\bibliography{References}

\end{document}